\documentclass[
  reprint,
  prd,
  superscriptaddress,
  nofootinbib
]{revtex4-2}

\usepackage{graphicx}
\usepackage{amsmath}
\usepackage{multirow}
\usepackage{caption}

\begin{document}

\title{Magnetic moments of strange hidden-bottom pentaquarks and the role of spin-flavor correlations}

\author{Pallavi Gupta}
\email{pallavi.gupta@thapar.edu} 
\affiliation{Department of Physics and Materials Science,
Thapar Institute of Engineering and Technology, Patiala, India}
\author{Vikash Kumar Garg}
\affiliation{Department of Computer Science and Engineering, Sant Longowal Institute of Engineering and Technology, Longowal, India }
\date{\today}
\begin{abstract}
We investigate the magnetic moments of strange hidden-bottom pentaquark states within the constituent quark model, considering both molecular and compact configurations. The system with quark content $qqqb\bar{b}$ ($q=u,d,s$) is analyzed in three configurations: a baryon-meson molecular form $(\bar b q_1)(b q_2 q_3)$, a diquark-diquark-antiquark structure $(b q_1)(q_2 q_3)\bar b$, and a diquark-triquark configuration $(b q_1)(\bar b q_2 q_3)$. Negative-parity states with $J^P = 1/2^-$, $3/2^-$, and $5/2^-$ are studied for strangeness $\mathcal{S}=-1,-2,-3$. For the dominant spin couplings, the two compact configurations yield identical or numerically very close magnetic moments. This indicates that the magnetic properties are governed
primarily by the global spin-flavor structure and
heavy-quark suppression effects rather than by the
specific clustering of quarks. A systematic suppression with increasing strangeness and a clear spin hierarchy are observed across all configurations. Due to the large bottom-quark mass, heavy-quark contributions are strongly suppressed, and the magnetic moments are dominated by light-strange spin correlations. These results provide useful theoretical benchmarks for future experimental and lattice studies of exotic multiquark states.
\end{abstract}

\maketitle

\section{Introduction}

The quark model introduced by Gell-Mann \cite{GellMann1964} Zweig \cite{Zweig1964} and Ne'eman \cite{Neeman1961} established SU(3) flavor symmetry as the organizing principle of hadron spectroscopy. Within this framework, conventional mesons and baryons are described as quark-antiquark $(q\bar{q})$ and three-quark $(qqq)$ configurations organized into SU(3) flavor multiplets. While this picture successfully organizes the conventional hadron spectrum, Quantum Chromodynamics does not restrict color-neutral states to these minimal configurations. Multiquark systems containing four $(qq\bar{q}\bar{q})$, five $(qqqq\bar{q})$, or more valence constituents are equally allowed by color confinement, motivating extensive theoretical and experimental investigations of exotic hadrons.

Clear experimental evidence for exotic hadrons emerged in the early 2000s. The observation of the $\chi_{c1}(3872)$ by the Belle Collaboration \cite{Choi2003} in 2003 initiated an intensive experimental and theoretical exploration of multiquark states. A decisive development occurred in 2015, when the LHCb Collaboration \cite{Aaij2015} reported the first experimental evidence of pentaquark states, $P_c(4380)^+$ and $P_c(4450)^+$, in the $\Lambda_b^0 \to J/\psi\, p\, K^-$ decay channel. These states were observed close to the $\Sigma_c \bar{D}$ and $\Sigma_c \bar{D}^{*}$ thresholds, suggesting a possible hadronic molecular interpretation $(qqqc\bar{c})$. Subsequent analyses revealed additional narrow states, including $P_c(4312)^+$, $P_c(4440)^+$, and $P_c(4457)^+$, as well as strange hidden-charm pentaquarks such as $P_{cs}(4459)^0$ and $P_{cs}(4338)^0$ \cite{Aaij2016, Aaij2019, Aaij2021, Adachi2025}. These observations firmly established pentaquarks as genuine exotic hadrons and triggered extensive theoretical investigations into their internal structure.

A variety of theoretical approaches have been developed to describe their spectroscopy and dynamics. Mass spectra and decay properties have been investigated using one-boson-exchange models \cite{Chen2019Decay, Wang2020}, QCD sum rules \cite{Zhang2019, Chen2019Interp}, effective field theories \cite{Lu2021, Peng2021}, and phenomenological quark models \cite{Meng2023, Giachino2023, Xiao2021, Ortega2023}. Despite their diversity, most interpretations can be broadly classified into three structural categories: hadronic molecular configurations \cite{He2019, Wang2020, Lu2021, Xiao2021}, diquark-diquark-antiquark models \cite{Anisovich2017, Shi2021, Mutuk2025Charm}, and diquark-triquark models \cite{Lebed2015, FernandezRamirez2019, Shimizu2019}. While different models account for selected experimental observations, masses and decay widths are insufficient to unambiguously identify the internal configuration. The dynamical structure of pentaquarks, therefore, remains an open question.

In this context, electromagnetic observables offer complementary insight. Magnetic moments, in particular, are directly sensitive to the spin–flavor structure and charge distribution of the constituents. As static electromagnetic observables, they provide complementary information beyond that contained in masses and decay widths and therefore offer valuable insight into the dominant internal correlations of exotic hadrons. They therefore offer complementary insight into the dominant internal correlations and provide information beyond that contained in masses and decay widths alone.
 Such quantities have been investigated extensively in the hidden-charm sector within both molecular and compact frameworks, including analyses of axial charges and radiative transitions \cite{Li2025, Lei2024, Li2024, Guo2024, Gao2022, Mutuk2025Charm}. 
In contrast, analogous investigations in the hidden-bottom sector remain limited. Most existing work focuses on spectroscopy, with electromagnetic properties explored primarily in the octet sector \cite{Mutuk2024BottomStrange, Mutuk2024Bottom}.

From heavy-quark symmetry considerations, bottom counterparts of the observed hidden-charm pentaquarks are expected from heavy-quark symmetry. Replacing a charm quark by a bottom quark preserves the underlying color structure while introducing a larger mass scale that modifies binding dynamics and suppresses heavy-quark contributions to electromagnetic observables. Experimentally, the LHCb Collaboration has reported exploratory analyses of structures containing a single bottom quark in weak decays mediated by the $b \to c\bar{c}s$ transition, observed in final states such as $J/\psi K\pi p$ and $J/\psi \phi p$ \cite{Aaij2018WeakbPc}. Although these studies do not establish fully hidden-bottom configurations, they indicate that the relevant production mechanisms are accessible, motivating dedicated searches for hidden-bottom pentaquarks.

In our previous work, we investigated the mass spectra of hidden-charm and hidden-bottom multiquark systems using the generalized Gürsey–Radicati mass formula and analyzed magnetic moments of hidden-charm pentaquarks within the molecular framework \cite{Gupta2026}. To our knowledge, a systematic study of magnetic moments in the strange decuplet hidden-bottom sector has not yet been reported. We compute the magnetic moments within three structural scenarios: a baryon–meson molecular form $(qqb)(q\bar{b})$, a diquark-diquark-antiquark model $((bq)\{q q\}\bar b)$, and a diquark-triquark configuration $((bq)(\bar b qq))$. In particular, we evaluate the magnetic moments of strange hidden-bottom pentaquark states with total spin-parity $J^P = \tfrac{1}{2}^-, \tfrac{3}{2}^-$, and $\tfrac{5}{2}^-$. This unified treatment enables a systematic comparison
of clustering schemes and clarifies the extent to which
spin-flavor correlations and heavy-quark suppression
govern the electromagnetic properties of hidden-bottom
pentaquarks.

The paper is organized as follows. In Sec.~II, we discuss the three structural models considered in this work and construct the corresponding wave functions of strange hidden-bottom pentaquark states. Section III presents the magnetic moment operator and analytical expressions. Numerical results are discussed in Section IV, followed by summary in Section V.

\section{Structural models}

In Quantum Chromodynamics, observable hadrons must be color singlets. Therefore, the internal structure of multiquark states is constrained by SU(3)$_c$ color symmetry. In this work, strange hidden-bottom pentaquark states are constructed by imposing the color-singlet condition and classifying the allowed configurations accordingly. To this end, it is instructive to examine
the color structures of the relevant quark clusters—mesons, baryons, diquarks, and
triquarks—from which physical pentaquark configurations can be formed.

\begin{table}[t]
\centering
\caption{Color structures of quark clusters and corresponding pentaquark models.}
\label{tab:combo}
\begin{tabular}{c c }
\hline\hline
Cluster type & Color composition  \\
\hline
Meson $(q\bar{q})$ 
& $3_c \otimes \bar{3}_c = 1_c \oplus 8_c$ 
 \\

Baryon $(qqq)$ 
& $3_c \otimes 3_c \otimes 3_c = 1_c \oplus 8_c \oplus 8_c \oplus 10_c$ 
 \\

Diquark $(qq)$ 
& $3_c \otimes 3_c = \bar{3}_c \oplus 6_c$ 
 \\

Triquark $(qq\bar{q})$ 
& $3_c \otimes 3_c \otimes \bar{3}_c = 3_{1c} \oplus 3_{2c} \oplus \bar{6}_c \oplus 15_c
$ 
 \\
\hline\hline
\end{tabular}
\end{table}

As summarized in Table \ref{tab:combo}, only specific color combinations of quark clusters can
lead to physical color-singlet pentaquark states. Enforcing this condition
naturally gives rise to three distinct structural scenarios of hidden-bottom
pentaquarks.

\noindent
\textbf{(i) Molecular configuration.}  
In this picture, the pentaquark is described as a loosely bound state of a
color-singlet meson and a color-singlet baryon. A quark-antiquark pair forming a
meson transforms under SU(3)$_c$ as
\begin{equation}
3_c \otimes \bar{3}_c = 1_c \oplus 8_c ,
\label{eq:color}
\end{equation}
while a three-quark baryon contains a color-singlet component. Since the direct
product of two color-singlet clusters satisfies
\begin{equation}
1_c \otimes 1_c = 1_c ,
\end{equation}
a color-singlet pentaquark can be formed. The corresponding quark configuration is
written as
\begin{equation}
(\bar b q_1)(b q_2 q_3)
\end{equation}

\noindent
\textbf{(ii) Diquark-diquark-antiquark configuration.}  
In this compact configuration, two quarks are first correlated into diquarks. The
color decomposition of a diquark is given by
\begin{equation}
3_c \otimes 3_c = \bar{3}_c \oplus 6_c ,
\end{equation}
where the color-antitriplet channel is attractive and therefore favored. Two
diquarks in the $\bar{3}_c$ representation can then combine as
\begin{equation}
\bar{3}_c \otimes \bar{3}_c = 3_c \oplus \bar{6}_c .
\end{equation}
The final coupling of the color-triplet component with the antiquark $\bar{3}_c$ representation proceeds according to the $SU(3)_c$ decomposition given in Eq.~(\ref{eq:color}), yielding a color-singlet pentaquark state. The corresponding
quark configuration is
\begin{equation}
(bq_1)(q_2q_3)\bar{b}.
\end{equation}

\noindent
\textbf{(iii) Diquark-triquark configuration.}  
In this model, the pentaquark is composed of a diquark and a triquark cluster. The
diquark is again assumed to form in the attractive color-antitriplet
representation. The triquark, consisting of two quarks and one antiquark, can form
a color-triplet state. The diquark–triquark coupling also follows the $SU(3)_c$ decomposition in Eq.~(\ref{eq:color}), ensuring overall color neutrality for the
pentaquark state. The corresponding quark configuration is written as
\begin{equation}
(bq_1)(\bar{b}q_2q_3).
\end{equation}

Although additional color-singlet combinations can be constructed in principle, some configurations are disfavored on dynamical grounds. In
particular, the $(\bar{b}b)$ cluster naturally forms a compact color-singlet
bottomonium state, while $(\bar{b}q_1)$ corresponds to a heavy-light meson. There
is no strong attractive interaction between a bottomonium state and a light baryon
$(q_1q_2q_3)$. Such configurations are therefore unlikely to form loosely bound molecular pentaquarks and are not considered in the present work.
\subsection*{Wave-function construction}

At the quark level, a hadronic system has four independent degrees of
freedom: color, spin, flavor, and space. Accordingly, the total wave function of
a pentaquark state can be written as a direct product of these components,
\begin{equation}
\Psi = \psi_{\text{flavor}}\,\psi_{\text{spin}}\,\psi_{\text{color}}\,\psi_{\text{space}} .
\end{equation}
Fermi-Dirac statistics require the total wave function to be antisymmetric under exchange of identical quarks in the light-quark subsystem.

In the present work, strange hidden-bottom pentaquark states are constructed
within the SU(3)$_f$ flavor symmetry framework. The flavor structure is governed
by the light-quark subsystem, while the bottom quark $b$ and the bottom
antiquark $\bar{b}$ are treated as flavor singlets. Depending on the strangeness
content, the pentaquark states considered here correspond to strangeness
$\mathcal{S}=-1$, $S=-2$, and $S=-3$.

We first consider the coupling of the two light quarks $q_2$ and $q_3$. In
SU(3)$_f$, their flavor decomposition is given by
\begin{equation}
3_f \otimes 3_f = \bar{3}_f \oplus 6_f .
\end{equation}
In the quark model, the antisymmetric $\bar{3}_f$ configuration corresponds to a scalar diquark ($S=0$), while the symmetric $6_f$ representation is associated with an axial-vector diquark ($S=1$).

Since the present analysis is restricted to states in the decuplet representation of SU(3)$_f$, the light-quark flavor wave function must be
fully symmetric. Consequently, only symmetric diquark configurations are
allowed. Throughout this work, a symmetric diquark is denoted by curly brackets
and defined as
\begin{equation}
\{q_2 q_3\} = \frac{1}{\sqrt{2}}\left(q_2 q_3 + q_3 q_2\right),
\end{equation}
which corresponds to an axial-vector diquark with spin $S=1$. Antisymmetric
scalar diquarks are therefore excluded from the present analysis.

If the $q_2 q_3$ pair is in the symmetric $6_f$ representation, it combines with
the remaining light quark $q_1$ according to
\begin{equation}
6_f \otimes 3_f = 10_f \oplus 8_f ,
\end{equation}
leading to decuplet and octet flavor multiplets. On the other hand, if the
$q_2 q_3$ pair is in the antisymmetric $\bar{3}_f$ representation, the flavor
coupling proceeds as
\begin{equation}
\bar{3}_f \otimes 3_f = 8_f \oplus 1_f ,
\end{equation}
resulting in octet and singlet flavor states.

After including the bottom quark $b$ and the bottom antiquark $\bar{b}$ and
employing the appropriate Clebsch-Gordan coefficients, the complete flavor wave
functions of hidden-bottom pentaquark states can be constructed. In the
molecular picture, this procedure yields the flavor wave functions
corresponding to the $(\bar{b}q_1)(bq_2q_3)$ configuration. The same construction
method can be straightforwardly extended to the diquark-diquark-antiquark
$(bq_1)(q_2q_3)\bar{b}$ and diquark-triquark $(bq_1)(\bar{b}q_2q_3)$
configurations.

For clarity and completeness, the explicit flavor-spin wave functions of the
strange hidden-bottom pentaquark states in all three structural models are
summarized in Table \ref{tab:wavefunctions}.

\begin{table}[htbp]
\caption{Wavefunctions of the strange hidden-bottom pentaquark in the decuplet multiplet across different models. The symbols I, $I_3$, Y, and $\mathcal{S}$ denotes Isospin, its third component, hypercharge, and strangeness, respectively. The bracket $\{q_1 q_2\}$ represents the symmetric diquark with $\{q_2 q_3\}= \sqrt{\frac{1}{2}}(q_2q_3+q_3 q_2) $.}
\label{tab:wavefunctions}
\begin{ruledtabular}
\begin{tabular}{c|c|c}
 $(I,I_3)$ & $(Y,\mathcal{S})$ & \textbf{Wavefunction} \\
\hline
\multicolumn{3}{c}{\textbf{Molecular Model $\bar b q_1 \{q_2q_3\}b$}} \\\hline\hline
 $(1,1)$ & \multirow{3}{*}{$(0,-1)$} & $\tfrac{1}{\sqrt3}\, \bar bs\,\{uu\}b + \sqrt{\tfrac{2}{3}}\, \bar b u\,\{us\}b$ \\
 $(1,0)$ &  &  $\frac{1}{\sqrt{3}}\left[\bar{b}s\{ud\}b+\bar{b}d\{us\}b
+\bar{b}u\{ds\}b\right]$\\
 $(1,-1)$  &  &  $\tfrac{1}{\sqrt3}\, \bar b s\,\{dd\}b
+ \sqrt{\tfrac{2}{3}}\, \bar b d\,\{ds\}b$ \\ 
\hline
$(\tfrac{1}{2},\tfrac{1}{2})$
& \multirow{2}{*}{$(-1,-2)$} & $\tfrac{1}{\sqrt3}\, \bar b u\,\{ss\}b
+ \sqrt{\tfrac{2}{3}}\, \bar b s\,\{us\}b$ \\[0pt]
\rule{0pt}{2.6ex}$(\tfrac{1}{2},-\tfrac{1}{2})$ &  & $\tfrac{1}{\sqrt3}\, \bar b d\,\{ss\}b
+ \sqrt{\tfrac{2}{3}}\, \bar b s\,\{ds\}b$ \\
 \hline
 $(0,0)$ & $(-2,-3)$ & $\bar b s\,\{ss\}b$   \\
\hline
\hline
\multicolumn{3}{c}{\textbf{Diquark Diquark Antiquark Model $(bq_3)\{q_1 q_2\}\bar b$}} \\\hline\hline
 $(1,1)$ & \multirow{3}{*}{$(0,-1)$} & $\tfrac{1}{\sqrt3}\, (bs)\{uu\}\bar b + \sqrt{\tfrac{2}{3}}\, (bu)\,\{us\}\bar b$ \\
 $(1,0)$ &  &  $\frac{1}{\sqrt{3}}\left[(bs)\{ud\}\bar b+ (bd)\{us\}\bar b
+(bu)\{ds\}\bar b\right]$\\
 $(1,-1)$  &  &  $\tfrac{1}{\sqrt3}\, (bs)\,\{dd\}\bar b
+ \sqrt{\tfrac{2}{3}}\, (bd)\{ds\}\bar b$ \\ 
\hline
$(\tfrac{1}{2},\tfrac{1}{2})$
& \multirow{2}{*}{$(-1,-2)$} & $\tfrac{1}{\sqrt3}\, (bu)\{ss\}\bar b
+ \sqrt{\tfrac{2}{3}}\, (bs)\{us\}\bar b$ \\[0pt]
\rule{0pt}{2.6ex}$(\tfrac{1}{2},-\tfrac{1}{2})$ &  & $\tfrac{1}{\sqrt3}\, (bd)\{ss\}\bar b
+ \sqrt{\tfrac{2}{3}}\, (bs)\{ds\}\bar b$ \\
 \hline
 $(0,0)$ & $(-2,-3)$ & $(bs)\{ss\}\bar b$   \\
\hline
\hline
\multicolumn{3}{c}{\textbf{Diquark-Triquark Model $(bq_3)(\bar b q_1 q_2)$}} \\\hline\hline
 $(1,1)$ & \multirow{3}{*}{$(0,-1)$} & $\tfrac{1}{\sqrt3}\, (bs)(\bar b\{uu\})+ \sqrt{\tfrac{2}{3}}\, (bu)\,(\bar b\{us\})$ \\
 $(1,0)$ &  &  $\frac{1}{\sqrt{3}}\left[(bs)(\bar b\{ud\})+ (bd)(\bar b\{us\})
+(bu)(\bar b\{ds\})\right]$\\
 $(1,-1)$  &  &  $\tfrac{1}{\sqrt3}\, (bs)\,(\bar b\{dd\})
+ \sqrt{\tfrac{2}{3}}\, (bd)(\bar b\{ds\})$ \\ 
\hline
$(\tfrac{1}{2},\tfrac{1}{2})$
& \multirow{2}{*}{$(-1,-2)$} & $\tfrac{1}{\sqrt3}\, (bu)(\bar b\{ss\})
+ \sqrt{\tfrac{2}{3}}\, (bs)(\bar b\{us\})$ \\[0pt]
\rule{0pt}{2.6ex}$(\tfrac{1}{2},-\tfrac{1}{2})$ &  & $\tfrac{1}{\sqrt3}\, (bd)(\bar b\{ss\})
+ \sqrt{\tfrac{2}{3}}\, (bs)(\bar b\{ds\})$ \\
 \hline
 $(0,0)$ & $(-2,-3)$ & $(bs)(\bar b\{ss\})$   \\
\end{tabular}
\end{ruledtabular}
\end{table}
\section{Magnetic Moment in Molecular Model}

For the molecular configuration $(\bar{b}q_1)(bq_2q_3)$, the magnetic moment arises from the spin contributions of the meson and baryon clusters. As the analysis is restricted to ground-state pentaquarks, the relative orbital angular momentum between the meson and baryon is taken to be zero, and orbital contributions are neglected.

The magnetic moment is a vector operator. Since the pentaquark states are taken with definite spin projection, we evaluate its expectation value along the quantization axis, i.e., the $z$-component, which corresponds to the observable magnetic moment.

The magnetic moment operator can therefore be written as
\begin{equation}
\hat{\mu} = \hat{\mu}_B + \hat{\mu}_M ,
\end{equation}
where $B$ and $M$ denote the baryon and meson clusters, respectively.

For the baryon cluster,
\begin{equation}
\hat{\mu}_B = \sum_{i=1}^{3} \mu_i g_i \hat{S}_{iz} ,
\end{equation}
with the sum running over its three constituent quarks. Here $\mu_i$ is the magnetic moment of the $i$-th quark, $g_i$ the corresponding Landé factor, and $\hat{S}_{iz}$ the spin operator.

Similarly, for the meson cluster,
\begin{equation}
\hat{\mu}_M = \sum_{i=1}^{2} \mu_i g_i \hat{S}_{iz} ,
\end{equation}
where the summation runs over the quark and antiquark constituents.
The pentaquark wavefunction $|\psi_{\text{Pentaquark}}\rangle$ follows the structure discussed in Section II. The magnetic moment is determined by the spin configuration, while the flavor content enters through the quark magnetic moments. It is then obtained as
\begin{equation}
\mu = \langle \psi_{\text{Pentaquark}} | (\hat{\mu}_B + \hat{\mu}_M)_z | \psi_{\text{Pentaquark}} \rangle .
\end{equation}
Carrying out the spin recoupling using standard Clebsch-Gordan coefficients, the general expression becomes

\begin{align}
\mu
&=
\sum_{S_z}
\langle S\, S_z \mid J\, J_z \rangle^{2}
\Bigg[
\sum_{S_{M_z}}
\langle S_B\, S_{B_z},\, S_M\, S_{M_z} \mid S\, S_z\rangle^{2}
\nonumber\\
&\quad\times
\Bigg(
S_{M_z}(\mu_{\bar b}+\mu_{q_1})
+
\sum_{S_{b_z}}
\langle S_b\, S_{b_z},\, S_D\, S_{D_z} \mid S_B\, S_{B_z}\rangle^{2}
\nonumber\\
&\qquad\quad\times
\big[
g\,\mu_b S_{b_z}
+
S_{D_z}(\mu_{q_2}+\mu_{q_3})
\big]
\Bigg)
\Bigg],
\end{align}

Here $S$, $S_B$, $S_M$, and $S_D$ denote the total spin of the pentaquark, baryon, meson, and the diquark inside the baryon, respectively, while $S_z$, $S_{B_z}$, $S_{M_z}$, and $S_{b_z}$ denote their corresponding spin projections. The analytical expressions for magnetic moments of all strange
hidden-bottom states in the molecular configuration for different combinations of $J^P =1/2^-,3/2^-,5/2^-$ are summarized
in Table~\ref{tab:magmom_mm}. 

\renewcommand{\arraystretch}{1.50}
\begin{table*}[htbp]
\caption{Magnetic moment of strange hidden bottom pentaquark states in the molecular configuration. The quantities in parentheses denote the isospin and its third component $(I, I_3)$. $J_{B}^{P_{B}}$, $J_{M}^{P_{M}}$, and $L^{P_{L}}$ correspond to the spin-parity quantum numbers of the baryon, meson, and relative orbital angular momentum, respectively. All magnetic moments are expressed in units of the proton magnetic moment.}
\label{tab:magmom_mm}
\begin{tabular}{c c  cc| cc |cc}
\hline
\hline
\multirow{2}{*}{$J^P$} &
& \multicolumn{2}{c}{\textbf{$P^+_{b\bar{b}s} (1,1)$}}
& \multicolumn{2}{c}{\textbf{$P^0_{b\bar{b}s} (1,0)$}}
& \multicolumn{2}{c}{\textbf{$P^-_{b\bar{b}s} (1,-1)$}}\\
\cline{3-8}
 &  \small $J_B^{P_B} \otimes J_M^{P_M} \otimes L^{P_L}$ & Expression & Value & Expression & Value & Expression & Value \\
\cline{1-8}
\multirow{3}{*}{$\tfrac12^-$}
& $\tfrac12^+\!\otimes0^-\!\otimes0^+$
& $\frac{1}{9}\left(8\,\mu_u + 4\,\mu_s - 3\,\mu_b\right)$
& 1.389
& $\frac{1}{9}\left(4\,\mu_u + 4\,\mu_d + 4\,\mu_s - 3\,\mu_b\right)$
& 0.170 &  $\frac{1}{9}\left(8\,\mu_d + 4\,\mu_s - 3\,\mu_b\right)$
& -1.050\\

& $\tfrac12^+\!\otimes1^-\!\otimes0^+$

& $\frac{1}{27}\left(4\,\mu_u + 2\,\mu_s - 15\,\mu_b\right)$ & 0.265
& $\frac{1}{27}\left(2\,\mu_u + 2\,\mu_d + 2\,\mu_s - 15\,\mu_b\right)$ & 0.062 & $\frac{1}{27}\left(4\,\mu_d + 2\,\mu_s - 15\,\mu_b\right)$ & -0.141\\

&  $\tfrac32^+\!\otimes1^-\!\otimes0^+$
&$\frac{1}{27}\left(14\,\mu_u + 7\,\mu_s +24\,\mu_b\right)$ & 0.738
&$\frac{1}{27}\left(7\,\mu_u + 7\,\mu_d + 7\,\mu_s + 24\,\mu_b\right)$& 0.026 & $\frac{1}{27}\left(14\,\mu_d + 7\,\mu_s +24\,\mu_b\right)$ & -0.685\\
\cline{1-8}

\multirow{3}{*}{$\tfrac32^-$}
&$\tfrac12^+\!\otimes1^-\!\otimes0^+$
 
& $\frac{1}{9}\left(14\,\mu_u + 7\,\mu_s - 12\,\mu_b\right)$ & 2.481
& $\frac{1}{9}\left(7\,\mu_u + 7\,\mu_d + 7\,\mu_s -12\,\mu_b\right)$ & 0.347 & $\frac{1}{9}\left(14\,\mu_d + 7\,\mu_s - 12\,\mu_b\right)$  & -1.786\\

& $\tfrac32^+\!\otimes0^-\!\otimes0^+$
 
& $\frac{1}{3}\left(4\,\mu_u + 2\,\mu_s + 3\,\mu_b\right)$ & 1.983
& $\frac{1}{3}\left(2\,\mu_u + 2\,\mu_d + 2\,\mu_s +3\,\mu_b\right)$ & 0.154 & $\frac{1}{3}\left(4\,\mu_d + 2\,\mu_s + 3\,\mu_b\right)$ & -1.674\\

& $\tfrac32^+\!\otimes1^-\!\otimes0^+$
& $\frac{1}{45}\left(56\,\mu_u + 28\,\mu_s +15\,\mu_b\right)$ & 1.891
& $\frac{1}{45}\left(28\,\mu_u + 28\,\mu_d + 28\,\mu_s +15\,\mu_b\right)$  & 0.184 & $\frac{1}{45}\left(56\,\mu_d + 28\,\mu_s +15\,\mu_b\right)$ & -1.522 \\
\cline{1-8} 
$\tfrac52^-$
&$\tfrac32^+\!\otimes1^-\!\otimes0^+$
& $2\,\mu_u + \mu_s$ & 3.075
& $\mu_u + \mu_d + \mu_s$ & 0.332 & $2\,\mu_d + \mu_s$ & -2.411\\
\hline \hline &
& \multicolumn{2}{c}{\textbf{$P^0_{b\bar{b}ss}(1/2,1/2)$}}
& \multicolumn{2}{c}{\textbf{$P^-_{b\bar{b}ss}(1/2,-1/2)$}}
& \multicolumn{2}{c}{\textbf{$P^-_{b\bar{b}sss}(0,0)$}}\\
\cline{3-8}
 &  & Expression & Value & Expression & Value & Expression & Value \\
\cline{1-8}

\multirow{3}{*}{$\tfrac12^-$}
& $\tfrac12^+\!\otimes0^-\!\otimes0^+$
& $\frac{1}{9}\left(4\,\mu_u + 8\,\mu_s - 3\,\mu_b\right)$
& 0.289
& $\frac{1}{9}\left(4\,\mu_d + 8\,\mu_s - 3\,\mu_b\right)$
& -0.930  &  $\frac{1}{3}\left(4\,\mu_s - \,\mu_b\right)$
& -0.811 \\

& $\tfrac12^+\!\otimes1^-\!\otimes0^+$

& $\frac{1}{27}\left(2\,\mu_u + 4\,\mu_s - 15\,\mu_b\right)$ & 0.082
& $\frac{1}{27}\left(2\,\mu_d + 4\,\mu_s - 15\,\mu_b\right)$& -0.121 & $\frac{1}{9}\left( 2\,\mu_s - 5\,\mu_b\right)$ & -0.102\\

&  $\tfrac32^+\!\otimes1^-\!\otimes0^+$
&$\frac{1}{27}\left(7\,\mu_u + 14\,\mu_s +24\,\mu_b\right)$ & 0.096
&$\frac{1}{27}\left(7\,\mu_d + 14\,\mu_s +24\,\mu_b\right)$& -0.615 & $\frac{1}{9}\left(7\,\mu_s +8\,\mu_b\right)$ & -0.545\\

\cline{1-8}

\multirow{3}{*}{$\tfrac32^-$}
&$\tfrac12^+\!\otimes1^-\!\otimes0^+$
 
& $\frac{1}{9}\left(7\,\mu_u + 14\,\mu_s - 12\,\mu_b\right)$ & 0.556
& $\frac{1}{9}\left(7\,\mu_d + 14\,\mu_s - 12\,\mu_b\right)$ & -1.577 & $\frac{1}{3}\left( 7\,\mu_s - 4\,\mu_b\right)$  & -1.369\\

& $\tfrac32^+\!\otimes0^-\!\otimes0^+$
 
& $\frac{1}{3}\left(2\,\mu_u + 4\,\mu_s + 3\,\mu_b\right)$ & 0.333
& $\frac{1}{3}\left(2\,\mu_d + 4\,\mu_s + 3\,\mu_b\right)$ & -1.496 & $2\,\mu_s + \mu_b$ & -1.317\\
& $\tfrac32^+\!\otimes1^-\!\otimes0^+$
& $\frac{1}{45}\left(28\,\mu_u + 56\,\mu_s +15\,\mu_b\right)$ & 0.351
& $\frac{1}{45}\left(28\,\mu_d + 56\,\mu_s +15\,\mu_b\right)$ & -1.356 & $\frac{1}{15}\left( 28\,\mu_s +5\,\mu_b\right)$ & -1.189 \\
\cline{1-8}
$\tfrac52^-$
&$\tfrac32^+\!\otimes1^-\!\otimes0^+$
& $\,\mu_u + 2\mu_s$ & 0.601
& $\,\mu_d + 2\mu_s$  & -2.143 & $3\mu_s$ & -1.875\\
\hline \hline
\end{tabular}
\end{table*}
As an illustrative example, consider a hidden-bottom pentaquark with
$J^{P} = \tfrac{1}{2}^{-}$ arising from the coupling
\[
J_B^{P_B} \otimes J_M^{P_M} \otimes L^{P_L}
= \tfrac{1}{2}^{+} \otimes 0^{-} \otimes 0^{+}.
\]

For the state with $S=-1$, $I=1$, $I_3=1$, and $Y=0$, the light-quark flavor
wave function is
\[
\frac{1}{\sqrt{3}}\,(\bar{b}s)\,\{uu\}b
+ \sqrt{\frac{2}{3}}\,(\bar{b}u)\,\{us\}b .
\]

Employing the magnetic-moment operator defined above we get 

\[
\begin{aligned}
\mu
&=
\left\langle
\tfrac{1}{2}\,\tfrac{1}{2}
\middle|
\tfrac{1}{2}\,\tfrac{1}{2}
\right\rangle^{2}
\Big[
\left\langle
\tfrac{1}{2}\,\tfrac{1}{2},\,0\,0
\middle|
\tfrac{1}{2}\,\tfrac{1}{2}
\right\rangle^{2}
\Big(
\left\langle
\tfrac{1}{2}\,\tfrac{1}{2},\,1\,0
\middle|
\tfrac{1}{2}\,\tfrac{1}{2}
\right\rangle^{2}
\mu_b
\\
&
\qquad
+
\left\langle
\tfrac{1}{2}\,-\tfrac{1}{2},\,1\,1
\middle|
\tfrac{1}{2}\,\tfrac{1}{2}
\right\rangle^{2}
\big(
-\mu_b
+\frac{2}{3}\mu_s
+\frac{4}{3}\mu_u
\big)
\Big)
\Big]
\\[6pt]
&=
\frac{1}{9}
\big(
-3\mu_b
+4\mu_s
+8\mu_u
\big).
\end{aligned}
\]
\section{Magnetic Moment in Diquark-Diquark-Antiquark Model}

In the configuration $(bq_1)(q_2q_3)\bar{b}$, the pentaquark is described as
two diquarks and a bottom antiquark. The magnetic moment operator
is written as
\begin{equation}
\hat{\mu} = \hat{\mu}_{D_1} + \hat{\mu}_{D_2} + \hat{\mu}_{\bar{b}},
\end{equation}
with $D_1=(bq_1)$ and $D_2=(q_2q_3)$.

For each axial-vector diquark,
\begin{equation}
\hat{\mu}_{D_i}
=
\sum_{k=1}^{2} \mu_k\, g_k\, \hat{S}_{kz} ,
\end{equation}
where the sum runs over its two constituent quarks. The antiquark contribution is
\begin{equation}
\hat{\mu}_{\bar{b}} = \mu_{\bar{b}}\, \hat{S}_{\bar{b}z} .
\end{equation}

The magnetic moment is obtained from
\begin{equation}
\mu = \langle \psi_{\text{Pentaquark}} | (\hat{\mu}_{D_1} + \hat{\mu}_{D_2} + \hat{\mu}_{\bar{b}})_z | \psi_{\text{Pentaquark}} \rangle.
\end{equation}

After performing the spin recoupling, the general expression becomes

\begin{align}
\mu
=&
\sum_{S_z}
\langle S S_z|J J_z\rangle^2
\Bigg[
\sum_{S_{\bar b z}}
\langle
S_{\bar b}S_{\bar b z},
S_{D_1D_2}S_{D_1D_2 z}
|S S_z
\rangle^2
\nonumber\\
&
\times
\Bigg(
S_{\bar b z}\mu_{\bar b}
+
\sum_{S_{D_1 z},S_{D_2 z}}
\nonumber\\
&
\qquad\times
\langle
S_{D_1}S_{D_1 z},
S_{D_2}S_{D_2 z}
|
S_{D_1D_2}S_{D_1D_2 z}
\rangle^2
\nonumber\\
&
\qquad\times
\Big[
S_{D_1 z}(\mu_b+\mu_{q_1})
+
S_{D_2 z}(\mu_{q_2}+\mu_{q_3})
\Big]
\Bigg)
\Bigg].
\end{align}
where $S_{D_1}$, $S_{D_2}$ and $S_{D_{1}D_2}$ denote the spins of the diquark $(bq_1)$,$(q_2q_3)$ and $(bq_1)(q_2q_3)$ respectively, while the corresponding spin projections are denoted by their z components. The corresponding results obtained in the
diquark–diquark–antiquark framework for the
$J^P = 1/2^-, 3/2^-, 5/2^-$ assignments are presented in
Table~\ref{tab:magmom_dd}.

\renewcommand{\arraystretch}{1.50}
\begin{table*}[htbp]
\caption{Magnetic moment in Diquark-Diquark Antiquark model for strange hidden bottom pentaquark states. The quantities in parentheses denote the isospin and its third component $(I, I_3)$. $J_{D_1}^{P_{D_1}}$, $J_{D_2}^{P_{D_2}}$, $J_{\bar{q}}^{P_{\bar{q}}}$, and $L^{P_{L}}$ denote the spin-parity quantum numbers of the clusters $(bq_1)$, $(q_2 q_3)$, the antiquark $\bar{b}$, and the relative orbital angular momentum, respectively. All magnetic moments are expressed in units of the proton magnetic moment.}
\label{tab:magmom_dd}
\begin{tabular}{c c cc| cc |cc}
\hline
\hline
\multirow{2}{*}{$J^P$} &
& \multicolumn{2}{c}{\textbf{$P^+_{b\bar{b}s} (1,1)$}}
& \multicolumn{2}{c}{\textbf{$P^0_{b\bar{b}s} (1,0)$}}
& \multicolumn{2}{c}{\textbf{$P^-_{b\bar{b}s} (1,-1)$}}\\
\cline{3-8}
 &  $\displaystyle J_{D_1}^{P_{D_1}} \otimes J_{D_2}^{P_{D_2}} \otimes J_{\bar q}^{P_{\bar q}} \otimes L^{P_L}$ & Expression & Value & Expression & Value & Expression & Value \\
\cline{1-8}
\multirow{3}{*}{$\tfrac12^-$}
& $ 0^{+} \otimes 1^{+} \otimes \tfrac{1}{2}^{-} \otimes 0^{+}$ & $\frac{1}{9}\left(  8\mu_u + 4\mu_s +3\mu_b \right)$ &1.344&$\frac{1}{9}\left( 4\mu_u + 4\mu_d + 4\mu_s+3\mu_b  \right)
$&0.125
&$\frac{1}{9}\left(  8\mu_d + 4\mu_s +3\mu_b  \right)$&-1.094\\
&$\displaystyle \left(1^{+} \otimes 1^{+}\right)_{0} \otimes \tfrac{1}{2}^{-} \otimes 0^{+}$&$-\mu_b
$&0.067&$-\mu_b
$&0.067&$-\mu_b
$&0.067\\
&$\displaystyle \left(1^{+} \otimes 1^{+}\right)_{1} \otimes \tfrac{1}{2}^{-} \otimes 0^{+}$&$\frac{1}{3}\left(2\mu_u + \mu_s + 2\mu_b   \right)
$&0.980&$\frac{1}{3}\left(  \mu_u + \mu_d + \mu_s +2\mu_b \right)
$&0.066&$\frac{1}{3}\left(2\mu_d + \mu_s + 2\mu_b   \right)
$&-0.848\\
\cline{1-8}

\multirow{3}{*}{$\tfrac32^-$}
&$\displaystyle \left(0^{+} \otimes 1^{+}\right) \otimes \tfrac{1}{2}^{-} \otimes 0^{+}$&$\frac{1}{3}\left( 4\mu_u + 2\mu_s -3\mu_b \right)
$&2.117&$\frac{1}{3}\left( 2\mu_u + 2\mu_d + 2\mu_s -3\mu_b  \right)
$&0.288&$\frac{1}{3}\left( 4\mu_d + 2\mu_s -3\mu_b \right)
$&-1.540\\
&$\displaystyle \left(1^{+} \otimes 1^{+}\right)_{1} \otimes \tfrac{1}{2}^{-} \otimes 0^{+}$&$\frac{1}{2}\left( 2\mu_u + \mu_s -\mu_b  \right)
$&1.571&$\frac{1}{2}\left( \mu_u + \mu_d + \mu_s -\mu_b \right)
$&0.199&$\frac{1}{2}\left( 2\mu_d + \mu_s -\mu_b  \right)
$&-1.172\\
&$\displaystyle \left(1^{+} \otimes 1^{+}\right)_{2} \otimes \tfrac{1}{2}^{-} \otimes 0^{+}$&$\frac{1}{10}\left( 18\mu_u + 9\mu_s+15\mu_b \right)
$&2.667&$\frac{1}{10}\left(  9\mu_u + 9\mu_d + 9\mu_s+15\mu_b  \right)
$&0.198&$\frac{1}{10}\left( 18\mu_d + 9\mu_s+15\mu_b \right)
$&-2.270\\
\cline{1-8}
$\tfrac52^-$
&$\displaystyle 1^{+} \otimes 1^{+} \otimes \tfrac{1}{2}^{-} \otimes 0^{+}$ &$2\mu_u + \mu_s
$&3.075&$\mu_u + \mu_d + \mu_s
$&0.332&$2\mu_d + \mu_s
$&-2.411\\
\hline \hline
&
& \multicolumn{2}{c}{\textbf{$P^0_{b\bar{b}ss}(1/2,1/2)$}}
& \multicolumn{2}{c}{\textbf{$P^-_{b\bar{b}ss}(1/2,-1/2)$}}
& \multicolumn{2}{c}{\textbf{$P^-_{b\bar{b}sss}(0,0)$}}\\
\cline{3-8}
 &  & Expression & Value & Expression & Value & Expression & Value \\
\cline{1-8}
\multirow{3}{*}{$\tfrac12^-$}
& $ 0^{+} \otimes 1^{+} \otimes \tfrac{1}{2}^{-} \otimes 0^{+}$ & $\frac{1}{9}\left(  4\mu_u + 8\mu_s +3\mu_b \right)$ &0.244&$\frac{1}{9}\left(  4\mu_d + 8\mu_s +3\mu_b \right)$ &-0.975
&$\frac{1}{3}\left(4\mu_s +\mu_b  \right)$&-0.855\\
&$\displaystyle \left(1^{+} \otimes 1^{+}\right)_{0} \otimes \tfrac{1}{2}^{-} \otimes 0^{+}$&$-\mu_b
$&0.067&$-\mu_b
$&0.067&$-\mu_b
$&0.067\\
&$\displaystyle \left(1^{+} \otimes 1^{+}\right)_{1} \otimes \tfrac{1}{2}^{-} \otimes 0^{+}$&$\frac{1}{3}\left(\mu_u + 2\mu_s + 2\mu_b   \right)
$&0.155&$\frac{1}{3}\left(\mu_d + 2\mu_s + 2\mu_b   \right)
$&-0.759&$\frac{1}{3}\left(3\mu_s + 2\mu_b   \right)
$&-0.700\\
\cline{1-8}

\multirow{3}{*}{$\tfrac32^-$}
&$\displaystyle \left(0^{+} \otimes 1^{+}\right) \otimes \tfrac{1}{2}^{-} \otimes 0^{+}$&$\frac{1}{3}\left( 2\mu_u + 4\mu_s -3\mu_b \right)
$&0.467&$\frac{1}{3}\left( 2\mu_d + 4\mu_s -3\mu_b \right)
$&-1.361&$ 2\mu_s -\mu_b 
$&-1.183\\
&$\displaystyle \left(1^{+} \otimes 1^{+}\right)_{1} \otimes \tfrac{1}{2}^{-} \otimes 0^{+}$&$\frac{1}{2}\left( \mu_u + 2\mu_s -\mu_b  \right)
$&0.333&$\frac{1}{2}\left( \mu_d + 2\mu_s -\mu_b  \right)
$&-1.038&$\frac{1}{2}\left( 3 \mu_s -\mu_b  \right)
$&-0.904\\
&$\displaystyle \left(1^{+} \otimes 1^{+}\right)_{2} \otimes \tfrac{1}{2}^{-} \otimes 0^{+}$&$\frac{1}{10}\left( 9\mu_u + 18\mu_s+15\mu_b \right)
$&0.439&$\frac{1}{10}\left( 9\mu_d + 18\mu_s+15\mu_b \right)
$&-2.029&$\frac{1}{10}\left( 27\mu_s+15\mu_b \right)
$&-1.788\\
\cline{1-8}
$\tfrac52^-$
&$\displaystyle 1^{+} \otimes 1^{+} \otimes \tfrac{1}{2}^{-} \otimes 0^{+}$ &$\mu_u + 2\mu_s
$&0.601&$\mu_d + 2 \mu_s
$&-2.143&$3\mu_s
$&-1.875\\
\hline \hline
\end{tabular}
\end{table*}

As an illustrative example, consider a hidden-bottom pentaquark
with $J^{P} = \tfrac{1}{2}^{-}$ arising from the coupling
\[
J_{D_1}^{P_{D_1}} \otimes J_{D_2}^{P_{D_2}}
\otimes J_{\bar q}^{P_{\bar q}} \otimes L^{P_L}
=
0^{+} \otimes 1^{+} \otimes \tfrac{1}{2}^{-} \otimes 0^{+}.
\]

For the state with $S=-1$, $I=1$, $I_3=1$, and $Y=0$, corresponding flavour wavefunction is \[\tfrac{1}{\sqrt3}\, (bs)\{uu\}\bar b + \sqrt{\tfrac{2}{3}}\, (bu)\,\{us\}\bar b\].
Using the general expression given in Eq.~(24), we get 

\begin{align}
\mu
=&
\left\langle
1\,1;\,
\tfrac{1}{2}\,-\tfrac{1}{2}
\mid
\tfrac{1}{2}\,\tfrac{1}{2}
\right\rangle^{2}
\nonumber\\
&
\times
\Bigg[
\mu_{\bar b}
+
\left\langle
0\,0;\,1\,1
\mid
1\,1
\right\rangle^{2}
\left(
\mu_u
+\frac{\mu_u}{3}
+\frac{2\mu_s}{3}
\right)
\Bigg]
\nonumber\\
&
+
\left\langle
1\,0;\,
\tfrac{1}{2}\,\tfrac{1}{2}
\mid
\tfrac{1}{2}\,\tfrac{1}{2}
\right\rangle^{2}
\Bigg[
\mu_{\bar b}
+
\left\langle
0\,0;\,1\,0
\mid
1\,0
\right\rangle^{2}
(0)
\Bigg].
\end{align}
Employing the Clebsch-Gordan coefficients for the above coupling
scheme and the relation $\mu_{\bar b} = -\,\mu_b $, the magnetic moment simplifies to
\[
\mu = \frac{1}{9}\left(  8\mu_u + 4\mu_s +3\mu_b \right)  .
\]
\section{Magnetic Moment in Diquark-Triquark Model}

In the diquark-triquark configuration $(bq_1)(\bar{b}q_2q_3)$,
the pentaquark is treated as a compact system composed of an
axial-vector diquark and a triquark cluster.

The magnetic moment operator can be written as
\begin{equation}
\mu = \langle \psi_{\text{Pentaquark}} | (\hat{\mu}_D + \hat{\mu}_T)_z | \psi_{\text{Pentaquark}} \rangle
\end{equation}
where $D=(bq_1)$ denotes the diquark and
$T=(\bar{b}q_2q_3)$ represents the triquark cluster.

For the axial-vector diquark,
\begin{equation}
\hat{\mu}_{D}
=
\sum_{i=1}^{2} \mu_i\, g_i\, \hat{S}_{iz} ,
\end{equation}
with the summation running over its two constituent quarks.

The triquark contribution arises from the bottom antiquark and
the two light quarks,
\begin{equation}
\hat{\mu}_{T}
=
\mu_{\bar{b}}\, \hat{S}_{\bar{b}z}
+
\sum_{j=2}^{3} \mu_{q_j}\, g_{q_j}\, \hat{S}_{q_j z}.
\end{equation}

The magnetic moment of the pentaquark state is obtained from
\begin{equation}
\mu
=
\langle \psi_{\text{Pentaquark}} |
(\hat{\mu}_{D} + \hat{\mu}_{T})
| \psi_{\text{Pentaquark}} \rangle .
\end{equation}

After performing the spin recoupling, the general expression becomes
\begin{align}
\mu
&= \sum_{S_z}
\langle S\, S_z \mid J\, J_z \rangle^{2}
\Bigg[
\sum_{S_{Dz},\, S_{Tz}}
\langle S_D\, S_{Dz},\, S_T\, S_{Tz} \mid S\, S_z \rangle^{2}
\nonumber\\
&\quad \times
\Bigg(
S_{Dz}(\mu_b+\mu_{q_1})
+
\sum_{S_{\bar b z},\, S_{D_i z}}
\langle S_{\bar b}\, S_{\bar b z},\, S_{D_i}\, S_{D_i z}
\mid S_T\, S_{Tz} \rangle^{2}
\nonumber\\
&\qquad \times
\big[
S_{\bar b z}\,\mu_{\bar b}
+
S_{D_i z}(\mu_{q_2}+\mu_{q_3})
\big]
\Bigg)
\Bigg].
\end{align}
where $S_D$, $S_T$ and $S_{D_i}$ denote the spins of the diquark $(bq_1)$, triquark $(\bar{b}q_2q_3)$ and light diquark $(q_2q_3)$ within triquark 
clusters, respectively, and the corresponding spon projections are denoted by their $z$ components. For completeness, the magnetic-moment formulas derived
within the diquark–triquark construction for
$J^P = 1/2^-, 3/2^-, 5/2^-$ are summarized in
Table~\ref{tab:magmom_dt}.
\begin{table*}[htbp]
\caption{Magnetic moment in Diquark-Triquark Model for strange hidden bottom pentaquark states. The quantities in parentheses denote the isospin and its third component $(I, I_3)$. $J_{D}^{P_{D}}$, $J_{T}^{P_{T}}$, and $L^{P_{L}}$ denote the spin-parity quantum numbers of the diquark, triquark, and the relative orbital angular momentum, respectively. All magnetic moments are given in units of the proton magnetic moment.}
\label{tab:magmom_dt}
\begin{tabular}{c c  cc| cc |cc}
\hline
\hline
\multirow{2}{*}{$J^P$} &
& \multicolumn{2}{c}{\textbf{$P^+_{b\bar{b}s} (1,1)$}}
& \multicolumn{2}{c}{\textbf{$P^0_{b\bar{b}s} (1,0)$}}
& \multicolumn{2}{c}{\textbf{$P^-_{b\bar{b}s} (1,-1)$}}\\
\cline{3-8}
 &  \small $J_D^{P_D} \otimes J_T^{P_T} \otimes L^{P_L}$ & Expression & Value & Expression & Value & Expression & Value \\
\cline{1-8}
\multirow{3}{*}{$\tfrac12^-$}
& $\tfrac12^-\!\otimes0^+\!\otimes0^+$
& $\frac{1}{9}\left(8\,\mu_u + 4\,\mu_s + 3\,\mu_b\right)$
& 1.344
& $\frac{1}{9}\left(4\,\mu_u + 4\,\mu_d + 4\,\mu_s + 3\,\mu_b\right)$
& 0.125 &  $\frac{1}{9}\left(8\,\mu_d + 4\,\mu_s + 3\,\mu_b\right)$
& -1.094\\

& $\tfrac12^-\!\otimes1^+\!\otimes0^+$

& $\frac{1}{27}\left(4\,\mu_u + 2\,\mu_s + 15\,\mu_b\right)$ & 0.190
& $\frac{1}{27}\left(2\,\mu_u + 2\,\mu_d + 2\,\mu_s + 15\,\mu_b\right)$ & -0.013 & $\frac{1}{27}\left(4\,\mu_d + 2\,\mu_s + 15\,\mu_b\right)$ & -0.216\\

&  $\tfrac32^-\!\otimes1^+\!\otimes0^+$
&$\frac{1}{27}\left(14\,\mu_u + 7\,\mu_s - 24\,\mu_b\right)$ & 0.857
&$\frac{1}{27}\left(7\,\mu_u + 7\,\mu_d + 7\,\mu_s - 24\,\mu_b\right)$& 0.146 & $\frac{1}{27}\left(14\,\mu_d + 7\,\mu_s - 24\,\mu_b\right)$ & -0.565\\
\cline{1-8}

\multirow{3}{*}{$\tfrac32^-$}
&$\tfrac12^-\!\otimes1^-\!\otimes0^+$
 
& $\frac{1}{9}\left(14\,\mu_u + 7\,\mu_s + 12\,\mu_b\right)$ & 2.302
& $\frac{1}{9}\left(7\,\mu_u + 7\,\mu_d + 7\,\mu_s +12\,\mu_b\right)$ & 0.169 & $\frac{1}{9}\left(14\,\mu_d + 7\,\mu_s + 12\,\mu_b\right)$  & -1.964\\

& $\tfrac32^-\!\otimes0^+\!\otimes0^+$
 
& $\frac{1}{3}\left(4\,\mu_u + 2\,\mu_s - 3\,\mu_b\right)$ & 2.117
& $\frac{1}{3}\left(2\,\mu_u + 2\,\mu_d + 2\,\mu_s -3\,\mu_b\right)$ & 0.288 & $\frac{1}{3}\left(4\,\mu_d + 2\,\mu_s - 3\,\mu_b\right)$ & -1.540\\

& $\tfrac32^-\!\otimes1^-\!\otimes0^+$
& $\frac{1}{45}\left(56\,\mu_u + 28\,\mu_s -15\,\mu_b\right)$ & 1.935
& $\frac{1}{45}\left(28\,\mu_u + 28\,\mu_d + 28\,\mu_s -15\,\mu_b\right)$  & 0.229 & $\frac{1}{45}\left(56\,\mu_d + 28\,\mu_s -15\,\mu_b\right)$ & -1.478 \\
\cline{1-8} 
$\tfrac52^-$
&$\tfrac32^-\!\otimes1^+\!\otimes0^+$
& $2\,\mu_u + \mu_s$ & 3.075
& $\mu_u + \mu_d + \mu_s$ & 0.332 & $2\,\mu_d + \mu_s$ & -2.411\\
\hline \hline &
& \multicolumn{2}{c}{\textbf{$P^0_{b\bar{b}ss}(1/2,1/2)$}}
& \multicolumn{2}{c}{\textbf{$P^-_{b\bar{b}ss}(1/2,-1/2)$}}
& \multicolumn{2}{c}{\textbf{$P^-_{b\bar{b}sss}(0,0)$}}\\
\cline{3-8}
 &  & Expression & Value & Expression & Value & Expression & Value \\
\cline{1-8}
\multirow{3}{*}{$\tfrac12^-$}
& $\tfrac12^-\!\otimes0^+\!\otimes0^+$
& $\frac{1}{9}\left(4\,\mu_u + 8\,\mu_s + 3\,\mu_b\right)$
& 0.244
& $\frac{1}{9}\left(4\,\mu_d + 8\,\mu_s + 3\,\mu_b\right)$
& -0.975  &  $\frac{1}{3}\left(4\,\mu_s + \,\mu_b\right)$
& -0.855 \\

& $\tfrac12^-\!\otimes1^+\!\otimes0^+$

& $\frac{1}{27}\left(2\,\mu_u + 4\,\mu_s + 15\,\mu_b\right)$ & 0.007
& $\frac{1}{27}\left(2\,\mu_d + 4\,\mu_s + 15\,\mu_b\right)$& -0.196 & $\frac{1}{9}\left( 2\,\mu_s + 5\,\mu_b\right)$ & -0.176\\

&  $\tfrac32^-\!\otimes1^+\!\otimes0^+$
&$\frac{1}{27}\left(7\,\mu_u + 14\,\mu_s -24\,\mu_b\right)$ & 0.215
&$\frac{1}{27}\left(7\,\mu_d + 14\,\mu_s -24\,\mu_b\right)$& -0.496 & $\frac{1}{9}\left(7\,\mu_s -8\,\mu_b\right)$ & -0.426\\

\cline{1-8}

\multirow{3}{*}{$\tfrac32^-$}
&$\tfrac12^-\!\otimes1^-\!\otimes0^+$
 
& $\frac{1}{9}\left(7\,\mu_u + 14\,\mu_s + 12\,\mu_b\right)$ & 0.378
& $\frac{1}{9}\left(7\,\mu_d + 14\,\mu_s + 12\,\mu_b\right)$ & -1.756 & $\frac{1}{3}\left( 7\,\mu_s + 4\,\mu_b\right)$  & -1.548\\

& $\tfrac32^-\!\otimes0^+\!\otimes0^+$
 
& $\frac{1}{3}\left(2\,\mu_u + 4\,\mu_s - 3\,\mu_b\right)$ & 0.467
& $\frac{1}{3}\left(2\,\mu_d + 4\,\mu_s - 3\,\mu_b\right)$ & -1.361 & $2\,\mu_s - \mu_b$ & -1.183\\
& $\tfrac32^-\!\otimes1^-\!\otimes0^+$
& $\frac{1}{45}\left(28\,\mu_u + 56\,\mu_s -15\,\mu_b\right)$ & 0.396
& $\frac{1}{45}\left(28\,\mu_d + 56\,\mu_s -15\,\mu_b\right)$ & -1.311 & $\frac{1}{15}\left( 28\,\mu_s -5\,\mu_b\right)$ & -1.144 \\
\cline{1-8}
$\tfrac52^-$
&$\tfrac32^-\!\otimes1^+\!\otimes0^+$
& $\,\mu_u + 2\mu_s$ & 0.601
& $\,\mu_d + 2\mu_s$  & -2.143 & $3\mu_s$ & -1.875\\
\hline \hline
\end{tabular}
\end{table*}
As an illustrative example, consider a strange hidden-bottom
pentaquark with $J^{P} = \tfrac{1}{2}^{-}$ arising from the coupling
\[
J_{D}^{P_{D}} \otimes J_{T}^{P_{T}} \otimes L^{P_L}
=
\tfrac{1}{2}^{-} \otimes 0^{+} \otimes 0^{+}.
\]

For the state with $S=-1$, $I=1$, $I_3=1$, and $Y=0$,
the corresponding flavor wave function is
\[
\tfrac{1}{\sqrt3}\, (bs)(\bar b\{uu\})+ \sqrt{\tfrac{2}{3}}\, (bu)\,(\bar b\{us\})
\]

Using the general expression derived above we get,
\begin{align}
\mu
&=
\left\langle \tfrac{1}{2}\, \tfrac{1}{2};\, 0\, 0 \mid \tfrac{1}{2}\, \tfrac{1}{2} \right\rangle^{2}
\Bigg[
\left\langle \tfrac{1}{2}\, \tfrac{1}{2};\, 1\, 0 \mid \tfrac{1}{2}\, \tfrac{1}{2} \right\rangle^{2}
\, \mu_{\bar b}
\Bigg]
\nonumber\\
&\quad+
\left\langle \tfrac{1}{2}\, -\tfrac{1}{2};\, 1\, 1 \mid \tfrac{1}{2}\, \tfrac{1}{2} \right\rangle^{2}
\Bigg[
-\mu_{\bar b}
+ \frac{4}{3}\mu_u
+ \frac{2}{3}\mu_s
\Bigg]
\nonumber\\[4pt]
&=
\frac{1}{9}
\left(
8\mu_u + 4\mu_s + 3\mu_b
\right).
\end{align}
\section{Numerical Analysis and Discussion}

In this section, we present a numerical analysis of the magnetic moments of strange hidden-bottom pentaquark states within the molecular and compact configurations considered in this work. The constituent quark masses adopted (in GeV) \cite{Mutuk2024BottomStrange} are
\begin{equation}
m_u = 0.338, \quad 
m_d = 0.350, \quad 
m_s = 0.500, \quad 
m_b = 4.67.
\end{equation}
The quark magnetic moments are evaluated using
\begin{equation}
\mu_q = \frac{e_q}{2 m_q},
\end{equation}
which yields the quark magnetic moments in units of the proton magnetic moment.
\begin{equation}
\mu_u = 1.851, \quad
\mu_d = -0.894, \quad
\mu_s = -0.626, \quad
\mu_b = -0.067.
\end{equation}
The very small magnitude of the bottom-quark magnetic moment immediately indicates that heavy-quark contributions are strongly suppressed in hidden-bottom systems.

A central result of our analysis is the near equivalence of
the two compact configurations, namely the diquark–diquark–
antiquark and diquark–triquark constructions. For the
dominant spin couplings and maximally aligned configurations,
both pictures lead to identical analytical expressions and
numerically indistinguishable magnetic moments. In particular,
the equivalence is exact for the maximally aligned
$J^P = \frac{5}{2}^-$ states and for the leading
$\frac{1}{2}^-$ configurations, while moderate differences
appear in certain $\frac{3}{2}^-$ and subleading
$\frac{1}{2}^-$ spin couplings. Nevertheless, the overall
magnetic properties remain largely governed by the total
spin–flavor configuration rather than by the specific
clustering of quarks into diquark substructures. In the
following discussion, we therefore treat them collectively
as a single compact description without loss of generality.

The observed similarity between different compact configurations carries an important physical implication. Since the magnetic moments are dominated by the contributions of the light and strange quarks, while the bottom-quark contribution is strongly suppressed, the resulting magnetic properties depend primarily on the overall spin-flavor structure of the pentaquark state. Consequently, configurations sharing the same spin-flavor content naturally lead to similar magnetic moments, despite differences in their internal clustering schemes. This behavior suggests that, in hidden-bottom systems, magnetic moments are more sensitive to global spin-flavor correlations than to the detailed arrangement of quarks within the state.

We now compare the compact and molecular configurations.
A clear trend emerges as the strangeness quantum number
varies from $\mathcal{S}=-1$ to $\mathcal{S}=-3$. For a fixed spin assignment
and electric charge, the magnitude of the magnetic moment
systematically decreases with increasing strange-quark
content. This behavior reflects the intrinsic hierarchy
$|\mu_u| > |\mu_s| > |\mu_b|$ within the constituent
quark model and follows directly from flavor composition:
increasing strangeness replaces light up quarks with
strange quarks, thereby reducing the overall magnetic
moment scale.

The $\mathcal{S}=-3$ sector is particularly instructive.
In the absence of up–down flavor asymmetry, only a
single isospin state remains, whose magnetic moment is
dominated by strange-quark contributions and exhibits
the strongest overall suppression.

For a fixed strangeness and electric charge sector,
the magnetic moments exhibit a robust spin hierarchy,
\begin{equation}
\mu\left(\frac{5}{2}^-\right) >
\mu\left(\frac{3}{2}^-\right) >
\mu\left(\frac{1}{2}^-\right).
\end{equation}
This ordering is preserved in both compact and molecular descriptions. It originates from the progressive alignment of light-quark spins
with the total angular momentum, leading to constructive interference
among individual magnetic contributions in higher-spin states. The persistence of this hierarchy across structural models confirms that magnetic moments are dominated by global spin-flavor correlations rather than by detailed clustering dynamics.

Within the $\mathcal{S}=-1$ and $\mathcal{S}=-2$ sectors, isospin multiplet splittings follow the ordering
\begin{equation}
\mu(P^+) > \mu(P^0) > \mu(P^-),
\end{equation}
reflecting the electric charge difference between up and down quarks. As the strange content increases, the relative contribution of up and down quarks becomes diluted, reducing the width of the multiplet. In the $\mathcal{S}=-3$ sector, where no up-down asymmetry is present, the multiplet structure disappears and only a single magnetic moment value remains.

The heavy bottom quark contributes only marginally to the total magnetic moment due to its large mass,
\begin{equation}
\mu_b \propto \frac{1}{m_b}.
\end{equation}
Numerically, $|\mu_b| \ll |\mu_u|, |\mu_s|$, which explains why variations in the internal arrangement of the $b\bar b$ pair produce only subleading effects. As a result, the magnetic structure of strange hidden-bottom pentaquarks is controlled predominantly by the light and strange quark sectors, while the heavy quark pair acts effectively as a spectator at leading order. This heavy-quark suppression explains the equivalence of compact configurations, the universality of the strangeness-driven suppression, and the weak sensitivity of magnetic moments to internal clustering.

\begin{figure}[t]
\centering
\includegraphics[width=1\linewidth]{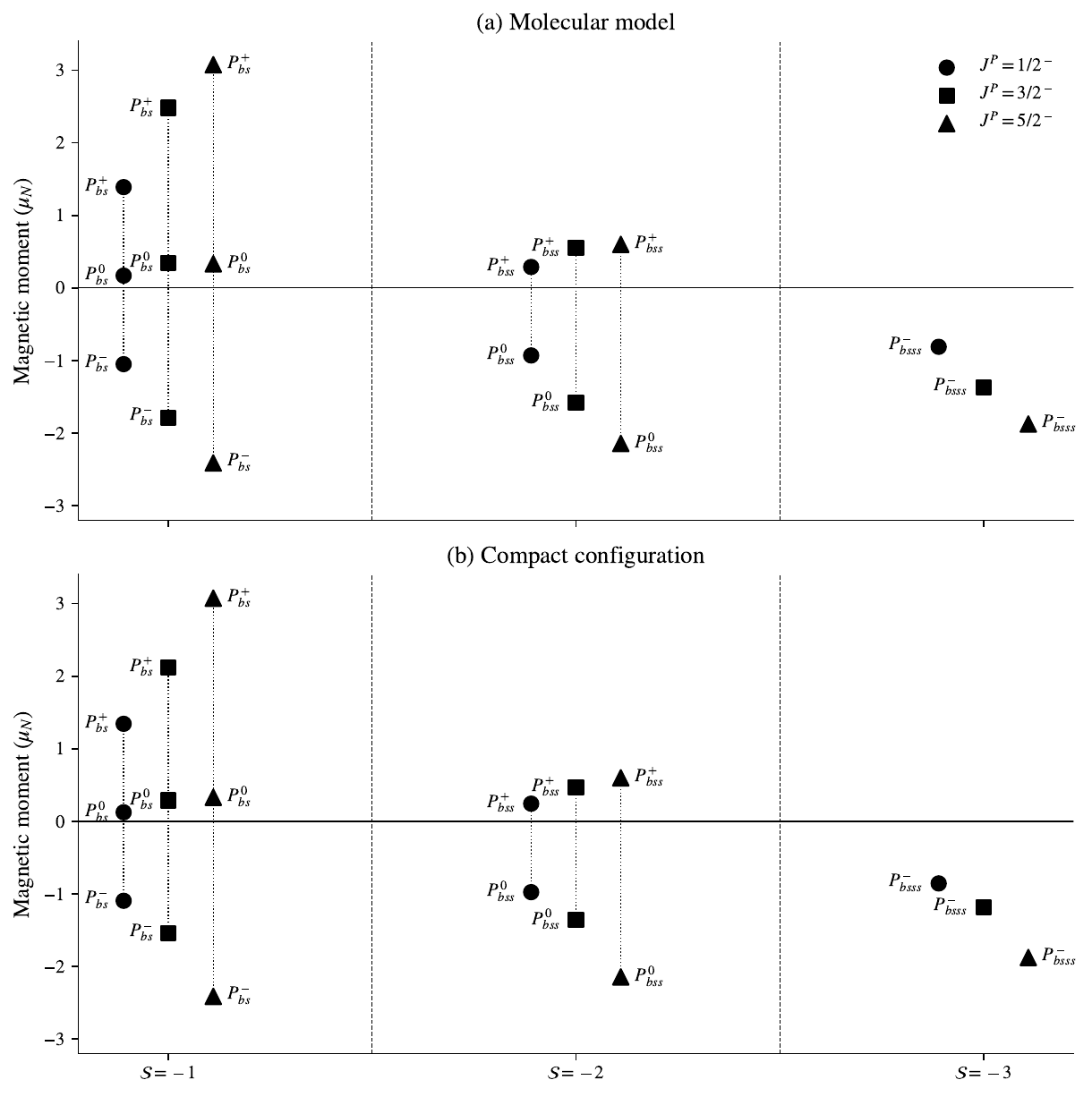}
\caption{
Magnetic moments of negative-parity strange hidden-bottom pentaquark states for $J^P = 1/2^-, 3/2^-, 5/2^-$. 
Panel (a) corresponds to the molecular configuration, while panel (b) shows the compact configuration. 
For clarity, only the leading spin configuration is shown.
}
\label{fig:magmom_strange}
\end{figure}

The numerical results are summarized in Fig.~\ref{fig:magmom_strange},
where the molecular configuration and the compact
diquark–diquark–antiquark scenario are compared directly.
For each $J^P = 1/2^-, 3/2^-, 5/2^-$, the figure displays
the leading (first) spin configuration in Tables~IV and~V,
including all available isospin members within the
$\mathcal{S}=-1$, $\mathcal{S}=-2$, and $\mathcal{S}=-3$ sectors.

The left panel illustrates the spin dependence within each
strangeness and charge sector, clearly displaying the
hierarchy $5/2^- > 3/2^- > 1/2^-$. The right panel highlights
the comparison between structural models and demonstrates
three robust features: (i) the systematic decrease of magnetic
moments with increasing strangeness, (ii) the persistence of
the spin ordering across all sectors, and (iii) the near
coincidence of the compact and molecular predictions for the
displayed configurations. The small differences observed
between the two scenarios arise from distinct spin-coupling
schemes and possible orbital contributions in the molecular
picture, but remain subleading due to the suppression of
heavy-quark contributions.

In the isospin-symmetric limit $\mu_u = \mu_d$, the multiplet splittings vanish entirely, while the monotonic suppression with increasing strangeness remains unchanged. This confirms that isospin splitting is electromagnetic in origin, whereas the global strangeness dependence is dictated by flavor composition.

Overall, the numerical analysis reveals that the magnetic moments of strange hidden-bottom pentaquarks are governed predominantly by light-strange spin correlations, with heavy-quark dynamics playing only a secondary role. It is instructive to compare the present results with other theoretical studies. In particular, Ref. \cite{Sharma2024} has investigated the magnetic moments of hidden-bottom pentaquarks using an effective mass scheme combined with a screened charge approach.
Despite the differences in methodology, our results exhibit similar qualitative features. In both approaches, the magnetic moments are primarily governed by the contributions of light quarks, while the heavy bottom quark and antiquark provide comparatively smaller contributions due to their large masses. Furthermore, both studies show consistent spin-dependent hierarchies and similar ordering patterns among different pentaquark states. 

The structural insensitivity of the results suggests that
future measurements of magnetic or transition magnetic
moments would provide valuable constraints on the
spin-flavor structure of these exotic states, while their
sensitivity to detailed internal clustering appears to be
limited. This behavior can be traced to the suppression of the
bottom-quark magnetic contribution and the dominance
of light-strange spin correlations in the magnetic
moment operator. In the present analysis, we have constructed pentaquark states using basis configurations with definite spin couplings and neglected possible mixing between these states. It is well known that, in a complete treatment, physical pentaquark states with definite masses can arise from mixing of different spin configurations. However, such mixing effects depend sensitively on the underlying interaction dynamics and model-dependent parameters, which are beyond the scope of the present work.
Since the magnetic moments are primarily governed by the dominant spin-flavor structure, the leading-order results obtained here are expected to capture the essential features of the pentaquark magnetic properties. The inclusion of mixing would lead to quantitative modifications but is not expected to alter the qualitative trends observed in this study. A systematic investigation of configuration mixing effects is left for future work. The present calculations are performed within the constituent quark model, which provides a phenomenological description of the dominant spin-flavor structure of multiquark systems. While quantitative predictions may depend on the theoretical framework employed, the qualitative features identified here, including heavy-quark suppression, strangeness dependence, and the spin hierarchy, are expected to be robust. Future studies using complementary approaches such as lattice QCD and effective field theories would be valuable for further assessing these results.

\section{Summary}

With the continued experimental discovery of exotic multiquark hadrons,
understanding their internal structure has become a central issue in hadron physics.
Magnetic moments are intrinsic observables that encode information
about the spin-flavor configuration of constituent quarks and therefore provide
valuable insight into the internal dynamics of exotic states.

In this work, we have systematically investigated the magnetic moments of strange
hidden-bottom pentaquark states within the constituent quark model framework.
Both molecular and compact configurations were considered, and the compact
scenario was analyzed in diquark-diquark-antiquark and diquark-triquark
constructions for the negative-parity states
$J^P = 1/2^-, 3/2^-, 5/2^-$ across the strangeness sectors
$\mathcal{S} = -1, -2, -3$.

An important conclusion of this work is that the two compact
descriptions lead to essentially the same magnetic behavior
in the physically relevant spin configurations. While certain
subleading couplings display moderate numerical deviations,
the overall pattern of magnetic moments is dictated primarily
by the total spin–flavor structure, exhibiting only limited sensitivity
to the particular internal clustering scheme. Owing to the large
bottom-quark mass, heavy-quark contributions are strongly suppressed, rendering
the magnetic structure predominantly sensitive to light-strange spin correlations. This heavy-quark  suppression explains the near equivalence of the compact configurations
and the limited sensitivity of the magnetic moments to detailed clustering effects.

A universal suppression with increasing strangeness and a robust spin hierarchy
are observed across all configurations. Although the molecular and compact
pictures differ in their internal organization, their predicted magnetic moments
remain numerically close, reflecting the structural insensitivity characteristic
of hidden-bottom systems.

The present results provide theoretical benchmarks for strange hidden-bottom
pentaquarks and clarify the dominant role of spin-flavor correlations in
determining their magnetic properties. A natural extension of this analysis
would be the investigation of open-heavy pentaquark configurations, where
flavor multiplet structures and spin couplings are substantially modified.
Such studies would further illuminate the interplay between heavy-quark symmetry,
spin-flavor correlations, and multiquark dynamics.

\end{document}